\documentclass{PoS}

\title{Curvature of the QCD phase transition line in a finite volume}

\ShortTitle{Curvature of the QCD phase transition line in a finite volume}

\author{\speaker{Bertram Klein}
\\
        Technische Universit\"at M\"unchen\\
        E-mail: \email{bklein@ph.tum.de}}

\author{Jens Braun\\
        Friedrich-Schiller-Universit\"at Jena\\
        E-mail: \email{j.braun@uni-jena.de}}
\author{Bernd-Jochen Schaefer\\
        Karl-Franzens-Universit\"at Graz\\
        E-mail: \email{bernd-jochen.schaefer@uni-graz.at}}

\abstract{The curvature which characterizes the QCD phase transition at finite temperature and small values of the chemical potential is accessible to lattice simulations. The results for this quantity which have been obtained by several different lattice simulation methods differ due to different numbers of flavors, different pion masses and different sizes of the simulation volume. In order to reconcile these results, it is important to investigate finite-volume effects on the curvature.  \\

We investigate the curvature of the chiral phase transition line at finite temperature and chemical potential in a finite volume. We use a phenomenological model for chiral symmetry breaking and apply non-perturbative functional renormalization group methods which account for critical long-range fluctuations at the phase transition.  \\

We find an intermediate volume region in which the curvature of the phase transition line is actually reduced relative to its infinite-volume value, provided periodic spatial boundary conditions are chosen for the quark fields. Size and location of this region depend on the value of the pion mass. Such an effect could account for differences in the curvature between lattice simulations in differently sized volumes and from functional methods in the infinite volume limit. We discuss implications of our results for the QCD phase diagram. }

\FullConference{The XXVIII International Symposium on Lattice Field Theory, Lattice2010\\
		June 14-19, 2010\\
		Villasimius, Italy}

\begin{document}

\section{Introduction}

There is considerable interest in determining the behavior of the QCD phase transition at finite temperature and chemical potential \cite{Fodor:2001au,Allton,deForcrand:2002,deForcrand:2003,deForcrand:2007,Karsch:2004}, see also \cite{lat2010curv} in these proceedings. Due to the complex phase of the fermion determinant, QCD at finite chemical potential cannot be simulated with standard Monte-Carlo methods on the lattice, but needs more sophisticated approaches, such as a Taylor-expansion in the chemical potential \cite{Allton,Karsch:2004} or simulation at imaginary chemical potential and analytic continuation \cite{deForcrand:2002,deForcrand:2003,deForcrand:2007}.

For small values of the chemical potential, the shape of the QCD phase transition line can be characterized by its curvature, which is accessible to the different simulation methods, as well as functional methods for QCD \cite{Braun:2009,Braun:2009gm} and for models, see e.g  \cite{Schaefer,Fukushima,Weise,Schaefer:2007pw,Skokov:2010wb,Herbst:2010rf,Skokov:2010uh,Karsch:2010hm,Cristoforetti:2010sn,Palhares:2009tf}. The curvature appears as the first expansion coefficient of the phase transition temperature, considered as a function of the quark chemical potential $\mu$. 

Due to calculations with different values of the pion mass and different numbers of flavors, the results for the curvature differ considerably between the different approaches \cite{deForcrand:2007,Karsch:2004}. An additional consideration are also the different sizes of the simulation volumes which may also affect the behavior of the transition. It is these effects that we wish to investigate in the present work.

When we put a system with fermions into a finite box, due to the anti-commutation relations the boundary conditions in the Euclidean time direction are required to be anti-periodic. In the spatial directions, however, we have a choice between periodic and anti-periodic boundary conditions. In many lattice QCD simulations, periodic boundary conditions are chosen in order to minimize finite-volume effects. In a model calculation, we have indeed found that that this choice minimizes finite-volume effects \cite{Braun:2005}.

In a small volume, the choice of boundary conditions can strongly influence the behavior of the system. 
The choice of boundary conditions affects the behavior of the pion mass in finite volume \cite{Braun:2005} as well as the chiral phase transition temperature \cite{Braun:2006}.
For periodic spatial boundary conditions, the presence of a quark zero-momentum mode leads to an enhancement of the chiral condensate.  In the limit of very small volume, the chiral condensate vanishes and chiral symmetry is restored, regardless of the boundary conditions.

In the framework of a model for chiral symmetry breaking, it follows from this observation that the chiral phase transition line at finite chemical potential ought to be affected by a finite volume: The presence of a chiral condensate leads to a finite mass for the constituent quarks in such a model. An enhancement of the chiral condensate translates directly into an increased constituent quark mass. More massive quarks then require a larger energy for their creation and a larger quark chemical potential is necessary to increase the quark density, compared to a situation with light constituent quarks. We expect that the system becomes therefore less sensitive to a change in the chemical potential, and consequently the phase transition line becomes flatter and the curvature decreases, in some intermediate volume region where this effect is present. 

We have investigated this hypothesis in the framework of a chiral model for values of the pion mass and of the volume that are relevant for current lattice simulations. We use a non-perturbative functional renormalization group method to correctly include Goldstone mode effects and critical fluctuations.

\section{Model and Method}
We use for our investigation the quark-meson model, a model for chiral symmetry breaking without any gluonic degrees of freedom, with two flavors of quarks. At an ultraviolet (UV) scale $\Lambda$, the model in Euclidean space-time is defined by the bare action
\begin{eqnarray} 
  \Gamma_{\Lambda}[\bar q,q,\phi]&=& \int \mathrm{d}^{4}x \Big\{
  \bar{q} \left({\partial}\!\!\!\!\!\slash + 
  h(\sigma+\mathrm{i}\vec{\tau}\cdot\vec{\pi}\gamma_{5})\right)q
  +\frac{1}{2}(\partial_{\mu}\phi)^{2}+U_{\Lambda}(\phi^2) - c \sigma \Big\},
\label{eq:QM}
\end{eqnarray}
where the four meson fields are parameterized as $\phi^{\mathrm{{T}}}=(\sigma,\vec{\pi})$.  
The meson potential with O(4) symmetry is given by 
\begin{equation}
  \label{eq:pot_UV} 
  U_\Lambda(\phi^{2}) =
  \frac{1}{2}m_\Lambda^{2}\phi^{2} +
  \frac{1}{4}\lambda_\Lambda(\phi^{2})^{2}
  \,,
\end{equation}
and thus characterized by the values of the two couplings $m_\Lambda$ and $\lambda_\Lambda$ at the UV scale. Explicit symmetry breaking due to a finite current quark mass $m_c$ is implemented via the term linear in $\sigma$ with coupling constant $c$
Spontaneous breaking of the chiral flavor symmetry is indicated by a finite expectation value for the first component of the meson field, $\langle \sigma \rangle \neq 0$. 

In order to investigate the behavior of the phase transition in this model in a finite volume, it is essential to take the critical long-range fluctuations into account. We employ a functional Renormalization Group (RG) method for this purpose, using an approach due to Wetterich \cite{Wetterich:1993}. The scale-dependent effective action $\Gamma_k$ for $\Lambda \ge k \ge 0$ satisfies the RG flow equation
\begin{equation}
k  \partial_k \Gamma_{k} = \frac{1}{2} \mathrm{STr}\left\{  \left[\Gamma _{k}^{(2)} + R_k\right]^{-1} \left(k\partial_k
      R_k\right) \right\}\, ,
\end{equation}
where $k$ is a sliding cutoff scale, and $R_k$ represents an infrared cutoff function, subject to certain constraints \cite{Wetterich:1993}. 
Here we use a smooth cutoff which can be directly related to an optimized regulator in the infinite-volume limit \cite{Litim,Litim:2001}. 
By solving the flow equation for the effective action in the above ansatz, we obtain in the limit $k \to 0$ the full effective action, including the effects of long-range fluctuations. The method can be adapted to a finite Euclidean space-time volume $L^3 \times 1/T$ \cite{Braun:2005,Braun:2006,Braun:2008sg,Braun:2010} by considering sums over discrete momentum modes instead of the continuous momentum integrations of the infinite-volume case. 
We solve the RG flow equation numerically by using an expansion of the local potential around its minimum; any additional momentum dependence of the couplings is neglected. 

\section{Results}
%
\begin{figure}\begin{center}
\includegraphics[scale=0.75]{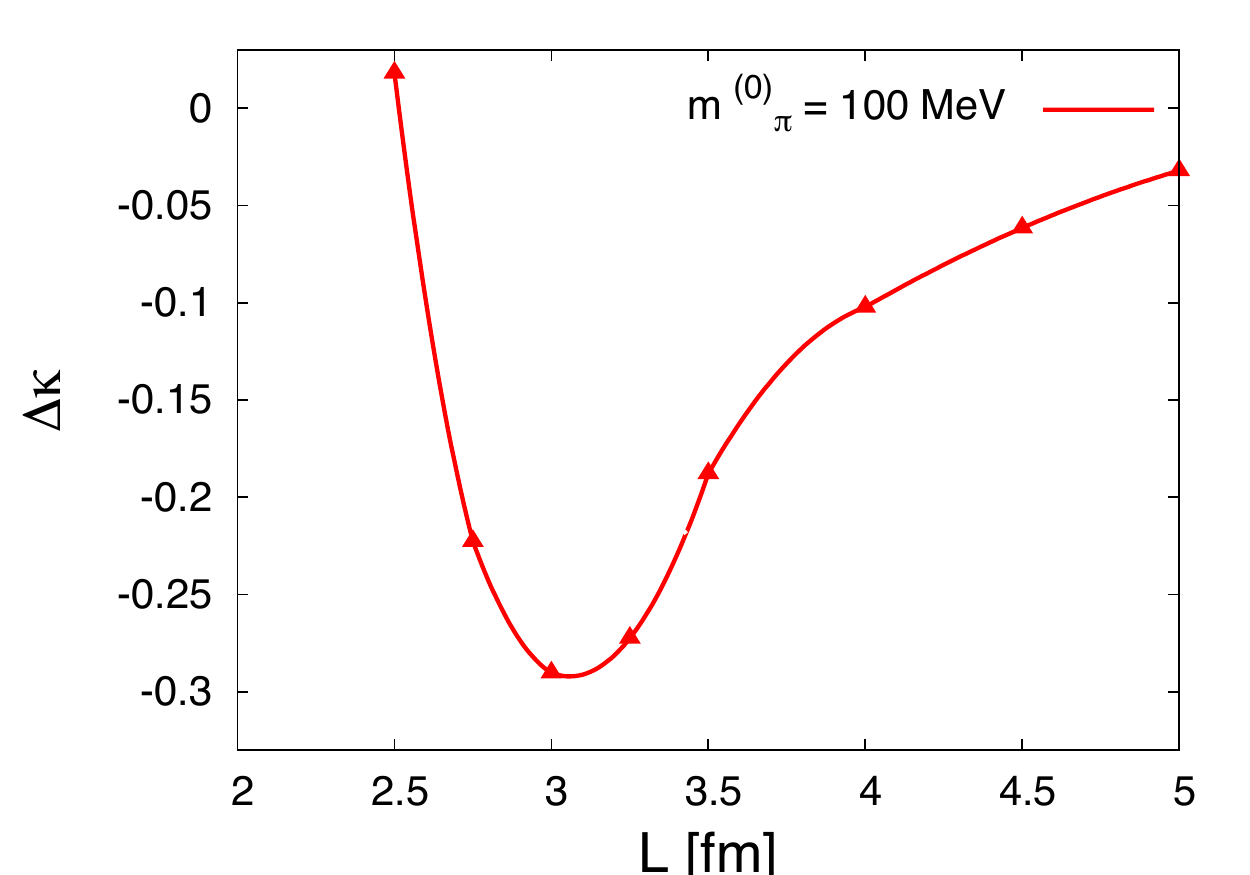}
\includegraphics[scale=0.75]{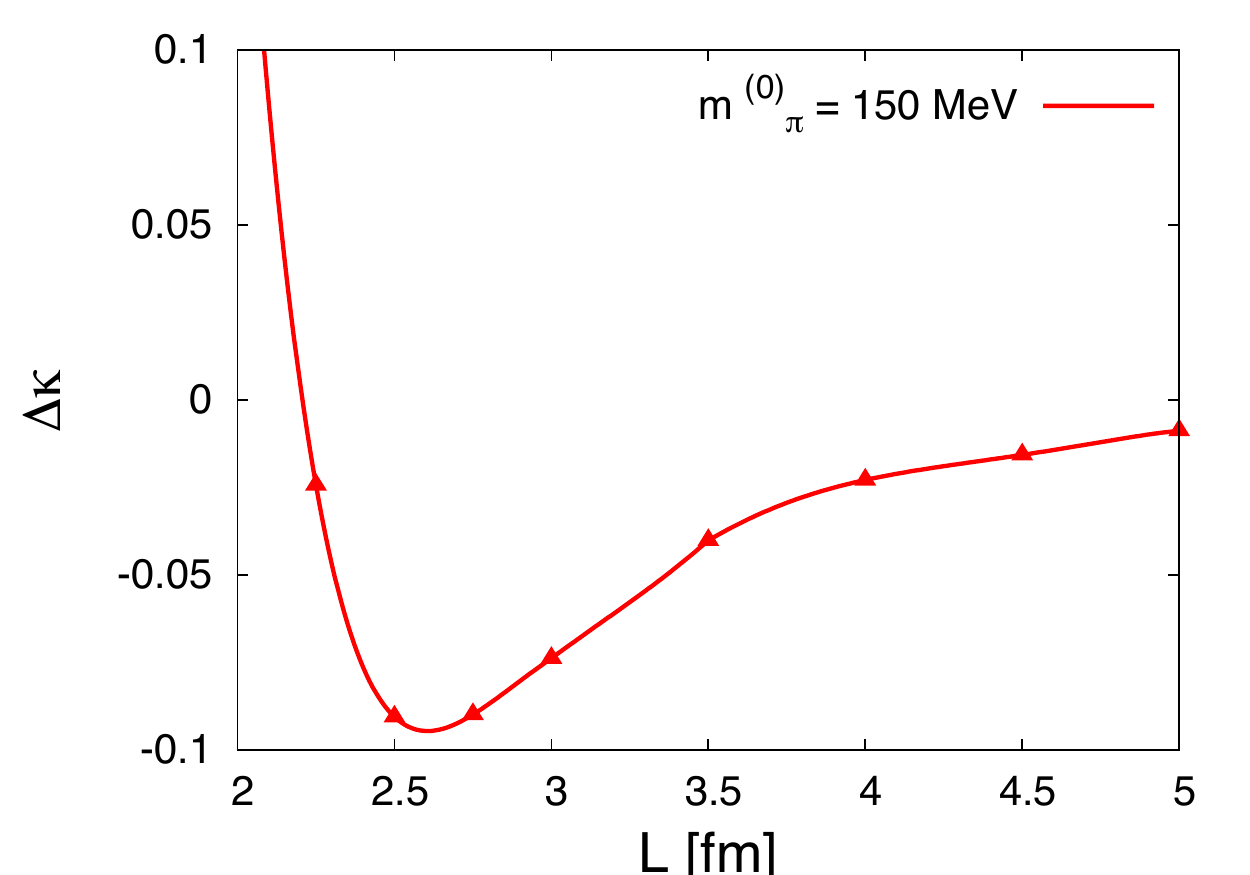}
\includegraphics[scale=0.75]{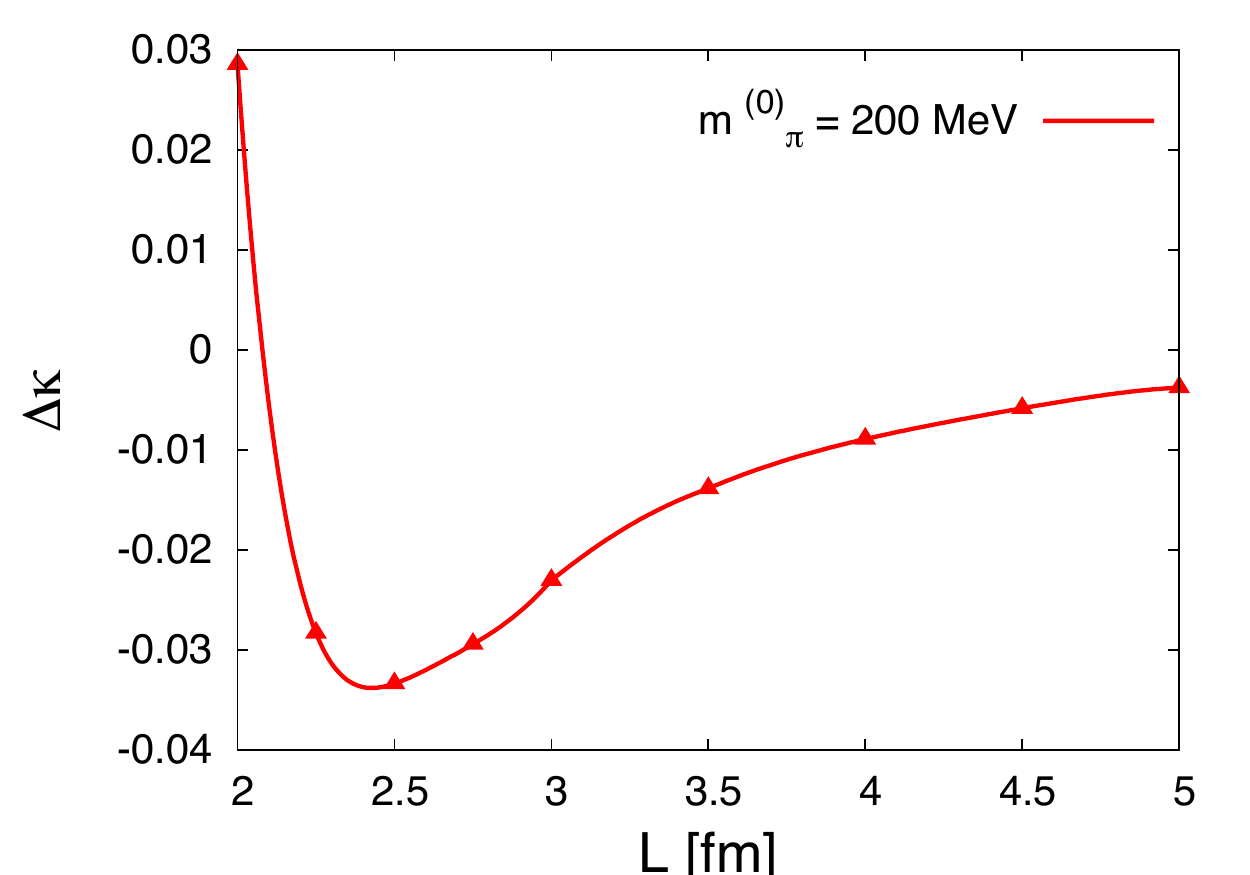}
\end{center}
\caption{Relative change  $\Delta \kappa$ of the curvature in a finite volume of box size $L$ compared to infinite volume for pion masses of $m_\pi^{(0)} =100$ MeV (top)  $m_\pi^{(0)} =150$ MeV (middle) and $m_\pi^{(0)} = 200$ MeV (bottom). Lines are meant to guide the eye. Results are obtained for periodic quark boundary conditions in the spatial direction.}
\label{fig:shift}
\end{figure}
We calculate the phase transition line at finite chemical potential and temperature for a variety of different values for the vacuum pion mass $m_\pi^{(0)}$ and different values for the box size $L$, including the limiting case $L \to \infty$. 
For the chiral phase transition, characterized by the behavior of the chiral condensate, we define the curvature $\kappa$ according to
\begin{equation}
  \frac{T_{\chi}(L,m_{\pi}^{(0)},\mu)}{T_{\chi}(L,m_{\pi}^{(0)},0)}=1-\kappa
  \left(\frac{\mu}{\pi T_{\chi}(L,m_{\pi}^{(0)},0)}\right)^2 +
  \dots\,,
\end{equation}
where $T_\chi(L, m_\pi^{(0)}, \mu)$ is the chiral phase transition temperature.
From the behavior of the crossover line, we extract the value of the curvature $\kappa$ for different volume sizes. In order to facilitate an easier comparison of the results, we introduce a relative shift for the curvature in finite volume
\begin{equation}
\Delta \kappa (L, m_\pi^{(0)}) =  \frac{\kappa(L,m_{\pi}^{(0)})- \kappa(\infty,m_{\pi}^{(0)})}
{\kappa(\infty,m_{\pi}^{(0)})}.
\end{equation}
The result for three exemplary values of the pion mass is shown in Fig.~\ref{fig:shift}. 

We find that the curvature \emph{decreases} in intermediate volume ranges, and only increases due to chiral symmetry restoration from finite-volume effects in very small volumes. In our model, for realistic values of the pion mass $\approx 150$ MeV, the decrease is as much as $10 \%$, but quickly becomes much smaller with increasing pion mass. The effects is specific to our choice of periodic quark boundary conditions. It is consistent with our expectation from the behavior of both the chiral condensate and the constituent quark mass in the model: due to the intermediate increase in the constituent quark mass, the system becomes less sensitive to an increase in the chemical potential, the phase transition line becomes flatter and the curvature decreases.

\section{Conclusions}
%
\begin{figure}
\centerline{\includegraphics[scale=0.8]{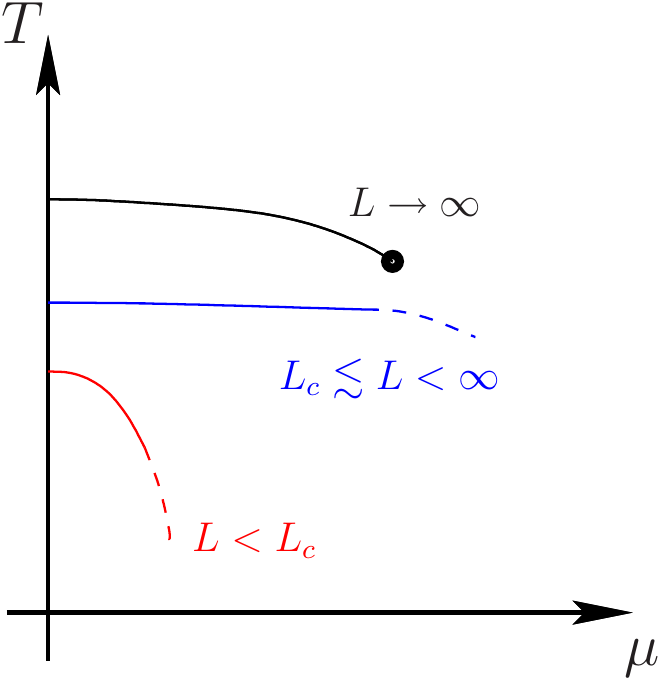}}
\caption{Schematic figure of the QCD phase diagram for finite chemical potential for different volume sizes $L$ at a given pion mass $m_\pi$. The solid lines symbolize the chiral crossover transition; the solid dot indicates the critical endpoint of a first-order line in the chiral phase diagram, as obtained in a study of the quark-meson model \cite{Schaefer}. In intermediate volume ranges, above a specific volume size $L_c$, the phase transition line is flatter than in infinite volume. For very small volumes, the curvature increases dramatically. The value of $L_c$ depends on the pion mass.}
\label{fig:FVpd}
\end{figure}
%
We have studied the effect of a finite volume on the chiral phase transition line in QCD at finite temperature and chemical potential. We used the quark-meson model, a model for chiral symmetry breaking that does not include gluonic and confinement effects. In order to take long-range fluctuations into account, which are of particular importance for studying the chiral phase transition, we have used a functional non-perturbative renormalization group method. 

For the choice of periodic spatial boundary conditions for the quark fields, we find that there are qualitatively clear effects on the curvature of the chiral phase transition line. In intermediate volume ranges -- depending on the exact values of the pion mass -- the phase transition line tends to flatten. Only for very small volumes, in which chiral symmetry is eventually restored, we do find an increase in the curvature.
This behavior is consistent with our expectations from investigations of the volume dependence of the pion mass and the chiral condensate \cite{Braun:2005}. The qualitative effect of this behavior on the phase diagram is sketched in Fig.~\ref{fig:FVpd}.

These observations have implications for the determination of the QCD phase diagram from lattice simulations as well. In general finite-volume effects will be stronger in the chiral model calculation, in particular in the absence of confinement, than in full QCD. In consequence, we do not expect to describe the finite-volume effects in QCD quantitatively. A more quantitative description could be achieved by including confinement effects via an effective potential for the Polyakov loop in finite volume. 

However, the finite-volume effects for the pion mass, which underlie our results for the phase boundary, have also been observed in quenched \cite{Guagnelli:2004ww} and in full \cite{Orth:2005kq} lattice QCD calculations and in a QCD calculation using Dyson-Schwinger equations \cite{Luecker:2009bs}. We therefore expect that such finite-volume effects are indeed present in lattice QCD simulations at finite chemical potential and may effect the shape of the phase diagram obtained from such simulations.

\section*{Acknowledgments}

B. K. acknowledges support by the DFG Research Cluster "Structure and Origin of the Universe". J. B. acknowledges support by the DFG Research Training Group "Quantum and Gravitational Fields" ({\mbox GRK 1523/1}).

\end{document}